\documentclass[10pt]{article}
\usepackage{graphicx} 
\usepackage{float}
\usepackage{amsthm}
\usepackage[hidelinks]{hyperref}
\usepackage{verbatim}
\usepackage{booktabs}  
\usepackage{xr-hyper}  
\usepackage{tabularx}
\usepackage{adjustbox}
\usepackage{makecell}
\usepackage{booktabs}
\usepackage{amsmath}
\usepackage{subcaption}
\usepackage{cite}

\usepackage[left=2.54cm, right=2.54cm]{geometry}
\newcolumntype{Y}{>{\raggedright\arraybackslash}X}

\usepackage{caption}
\usepackage[mathscr]{euscript} 
\usepackage{amsthm}
\usepackage{framed,multirow}

\usepackage{latexsym}
\usepackage{adjustbox}
\usepackage{tabu}
\usepackage{multicol}
\usepackage{soul}
\usepackage{blindtext}
\usepackage{multirow}
\usepackage{xr-hyper}
\makeatletter

\title{\texttt{MiNAA-WebApp}: A Web-Based Tool for the Visualization and Analysis of Microbiome Networks}

\usepackage{caption}
\usepackage{subcaption}
\usepackage{fancyhdr}
\usepackage{authblk} 
\usepackage[utf8]{inputenc}
\usepackage[T1]{fontenc}

\author[a,b,c]{\textbf{Qiyao Yang}}
\author[a]{\textbf{Rosa Aghdam}}
\author[a]{\textbf{Reed Nelson}}
\author[a,d,*]{\textbf{Claudia Sol\'{i}s-Lemus}}
\affil[a]{Wisconsin Institute for Discovery, University of Wisconsin-Madison, Madison, WI}
\affil[b]{Department of Computer Science, University of Wisconsin-Madison, Madison, WI}
\affil[c]{Department of Statistics, University of Wisconsin-Madison, Madison, WI}
\affil[d]{Department of Plant Pathology, University of Wisconsin-Madison, Madison, WI}
\affil[*]{corresponding author: solislemus@wisc.edu}

\begin{document}

\maketitle

\section{Summary}
Microbial networks, representing microbes as nodes and their interactions as edges, are crucial for understanding community dynamics in various environments. Analyzing microbiome networks is crucial for identifying keystone taxa that play central roles in maintaining microbial community structure and function, assessing how environmental changes such as pollution, climate shifts, or land use affect microbial dynamics, tracking disease progression by revealing alterations in microbial interactions over time, and predicting microbial community responses to interventions such as antibiotics, probiotics, or changes in diet and habitat. The complexity of microbial interactions necessitates the use of computational tools such as the \texttt{MiNAA-WebApp}, available at \url{https://minaa.wid.wisc.edu}, which enhances the accessibility of the Microbiome Network Alignment Algorithm (\texttt{MiNAA}). This tool allows researchers to align microbial networks and explore ecological relationships and community dynamics without extensive computational skills.
Originally, \texttt{MiNAA}'s command-line interface limited its usability for those without programming backgrounds. The web-based \texttt{MiNAA-WebApp} addresses this shortcoming by offering an intuitive interface with visualization tools, allowing easy exploration and analysis of microbial networks. The web app is designed for microbiome networks but also applicable to other biological networks, broadening its use in computational biology and making network-based research accessible to a wider audience.

\section{Statement of Need}
By aligning microbiome networks obtained under different settings (e.g. different fumigation treatments in soil), 
the Microbiome Network Alignment Algorithm (\texttt{MiNAA}) \cite{Nelson2024} method allows scientists to identify key microbial taxa that differs across treatments, as well as key taxa that serves similar function on the different communities.
The \texttt{MiNAA} algorithm builds on 
the GRAph ALigner (\texttt{GRAAL}) algorithm \cite{kuchaiev2010topological}, the Hungarian algorithm \cite{kuhn1955hungarian, Pilgrim1995} and the Graphlet Degree Vector (GDV) method \cite{prvzulj2007biological} to characterize nodes by local connectivity, capturing both direct connections and broader structural roles. 
The \texttt{MiNAA-WebApp} addresses the challenge of making advanced network analysis tools accessible to non-technical users to allow understanding of ecological dynamics. It provides a user-friendly, web-based platform that allows scientists to incorporate computational techniques into research and lowers barriers for exploring complex datasets. The application’s interactive visualizations and intuitive design support hands-on learning and make abstract concepts tangible, suitable for classroom use and independent study. It also supports interdisciplinary collaboration, enabling researchers from various fields to engage with the tool without programming expertise. The open-source nature of the \texttt{MiNAA-WebApp} allows for customization to meet specific teaching or research needs, democratizing access to advanced computational methods and promoting broader adoption across disciplines.

\section{Description of the web app}
The app is structured into four main tabs: Run, Visualization, Results, and About.

\underline{The Run Tab} facilitates data uploads and parameter adjustments for analyses with the \texttt{MiNAA} algorithm. Users select either a cost matrix or a similarity matrix through a toggle and can opt to generate visualizations with their analyses. The tab includes sliders and input fields for fine-tuning alignment parameters such as thresholds, node sizes, and degrees. A notable feature is the Color Options section, which allows users to customize the colors of nodes and edges in both networks (denoted G and H), and the connections between them, enhancing visual distinction. Adjustments can be viewed in real-time next to parameter settings, with additional options to refine the Size and Degree Parameters for clearer visualizations.
The Size Aligned parameter ($\geq 1$) determines the size of aligned nodes in the visualization. A larger value increases the size of aligned nodes, making them more visually prominent. The Degree Align parameter ($\geq 0$) sets the minimum degree a node must have to be displayed in the graph. Nodes with fewer connections than this threshold will be hidden. A higher Degree Align value filters out less connected nodes, reducing visual clutter and making it easier to focus on structurally significant elements. The Threshold Align parameter (range: $(0, 1]$) defines the minimum similarity required for nodes to be considered aligned. A higher threshold enforces stricter alignment, ensuring that only highly similar nodes are matched, while a lower threshold allows for more flexible alignments, which can be helpful when dealing with noisy or incomplete data.
Graphical outputs can be exported in common formats like JPEG or PNG for easy inclusion in reports and presentations. The Run Tab also provides a Metrics Summary to assess the alignment's effectiveness and quality, supporting comprehensive analysis and interpretation of complex networks. In addition to visual outputs, the Run Tab provides a Metrics Summary to help users evaluate the performance and quality of the network alignment.
To evaluate the structural alignment between two networks, we introduce two metrics: the Percentage of Aligned Edge Pairs and the Adjusted Number of Aligned Edge Pairs. The \textit{Percentage of Aligned Edge Pairs} quantifies the proportion of successfully aligned edges relative to the smaller network and is calculated as:
$
\text{Percentage of Aligned Edge Pairs} = \left( \frac{\text{number of aligned edge pairs}}{\min(\text{number of nodes in } G, \text{number of nodes in } H)} \right) \times 100.
$
This metric ensures that alignment success is normalized against the smaller network, allowing for fair comparisons across different datasets. For instance, if Network $G$ contains 50 nodes, Network $H$ contains 60 nodes, and 20 edges are aligned, the percentage is computed as:
$
\left( \frac{20}{50} \right) \times 100 = 40\%.$
This indicates that 40\% of the edges in the smaller network are successfully aligned.
To further refine alignment evaluation, we introduce the \textit{Adjusted Number of Aligned Edge Pairs}, which accounts for alignment cost and quality. It is computed as:
$
\text{Adjusted Number of Aligned Edge Pairs} = \text{number of aligned edge pairs} - \frac{\text{total cost}}{\text{number of aligned edge pairs}}.
$
Here, the \textit{total cost} represents the cumulative alignment cost derived from the first row of the alignment list. For example, if there are 20 aligned edge pairs and the total alignment cost is 5, the adjusted number is calculated as:
$
 20 - \frac{5}{20}= 19.75.
$
Together, these two measures provide a comprehensive evaluation of network alignment, balancing both quantity and quality to assess structural similarity effectively.

\begin{figure}[H] 
    \centering \includegraphics[width=0.7\textwidth]{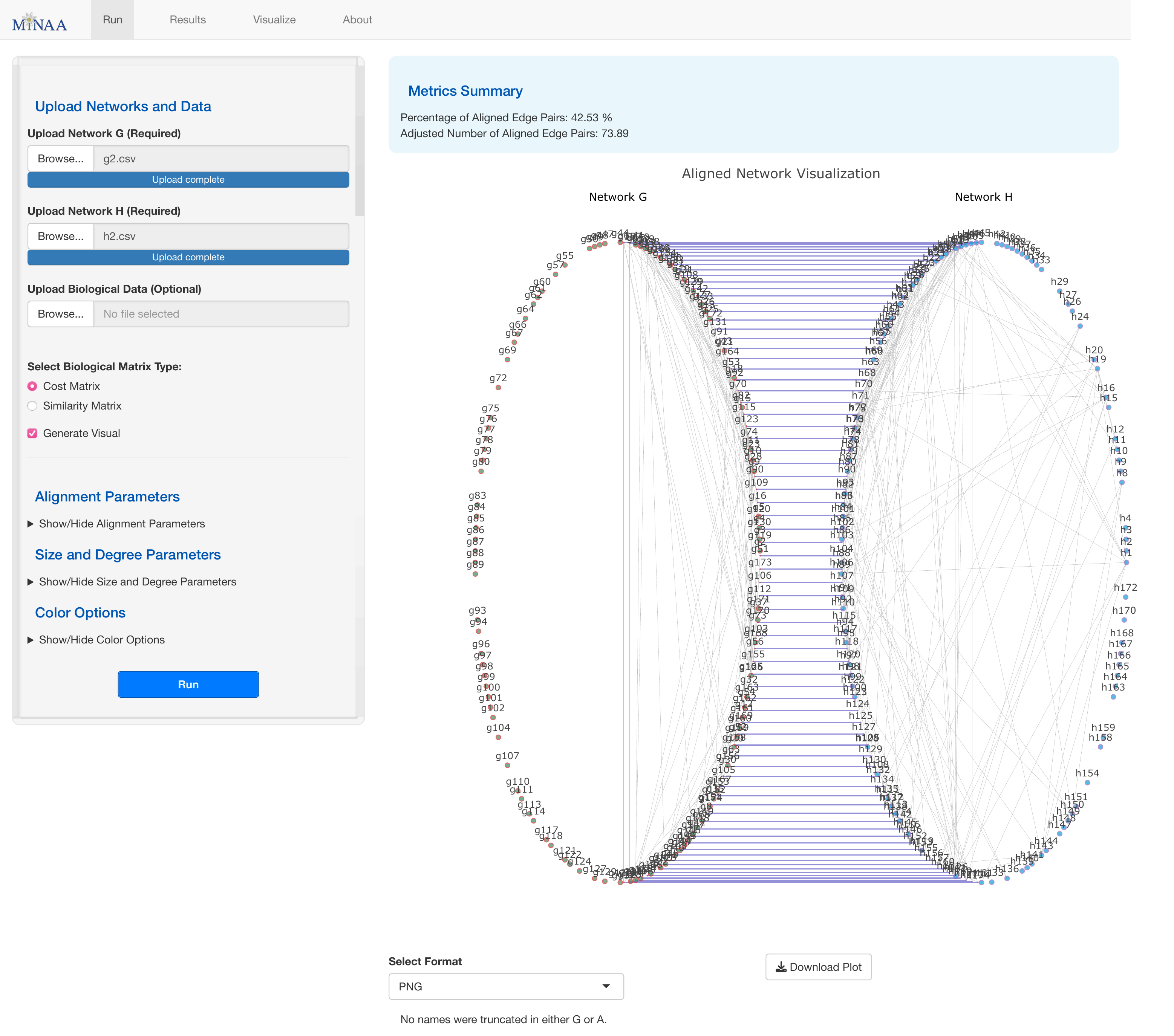}    \caption{Visualization of the \texttt{MiNAA-WebApp} interface, highlighting network alignment features and customizable parameter settings}
    \label{fig:miNAA-webapp}
\end{figure}

\underline{The Results Tab} allows users to access, manage, and interpret the outputs generated by the \texttt{MiNAA} algorithm. This tab provides a seamless interface for downloading result files, filtering outputs based on alignment thresholds, and understanding the key metrics and file structures associated with the analysis. Users begin by selecting the desired result folder, which contains all the output files generated from the alignment analysis. The Select and Manage Files section enables users to choose specific files for download through a dropdown menu, with a button to download the selected files efficiently. This ensures that users can quickly access the results they need for further analysis or documentation. The tab also provides options for refining alignment outputs by adjusting the alignment threshold. This Alignment Parameters section allows users to filter results dynamically, focusing on the most relevant alignments by modifying the minimum alignment score included in the outputs. A higher threshold (closer to 1) enforces stricter alignment filtering, ensuring that only high-confidence alignments are displayed. A lower threshold (closer to 0) includes more alignments, allowing for a broader but potentially noisier set of results. To ensure clarity and usability, the Metadata Information section offers detailed descriptions of the various output files generated by the analysis. These include the
Processed Data Files (g.csv and h.csv) which contain the processed data for networks G and H; the Graphlet Degree Vectors (g\_gdvs.csv and h\_gdvs.csv) which represent graphlet-based structural information (nodes signatures)  of networks G and H; the Log File (log.txt), providing a comprehensive record of the analysis, including key alignment steps and parameters used, and the Topological and Biological Cost Matrices (top\_costs.csv and bio\_costs.csv) summarize the topological and biological alignment costs, with the latter generated only when a biological similarity matrix is provided.

\underline{The Visualize Tab} enables users to upload data and interactively customize visualizations of network alignment results  generated by the \texttt{MiNAA} algorithm. This tab requires users to run \texttt{MiNAA} beforehand to generate the necessary alignment matrix. A clear message is displayed at the top of the tab, reminding users to complete this prerequisite step. Users can upload their processed network files (G and H) and the alignment matrix via dedicated file upload fields, which are marked as required. Once the files are uploaded, users can adjust parameters for size, alignment thresholds, and degree values, using intuitive sliders and input fields to refine the visualization. A Color Options section enables further customization, allowing users to set specific colors for nodes, edges, and alignment lines, making it easy to distinguish elements within the graph. The Visualize Tab also provides a real-time preview of the network alignment, allowing users to observe the effects of their parameter adjustments dynamically. Users can export the final graph in JPEG or PNG format for inclusion in reports, presentations, or other documentation. With its seamless integration of file uploads, parameter adjustments, and real-time visual feedback, the Visualize Tab complements the workflow established in the Run Tab. By ensuring that users have pre-processed their data with \texttt{MiNAA}, this tab streamlines the exploration of alignment results, providing flexibility and clarity for both research and educational applications.

\underline{The About Tab} provides an overview of the \texttt{MiNAA-WebApp}, including its purpose and functionality. The web page and accompanying web app are open-source, with the code available in the GitHub repository: \url{https://github.com/solislemuslab/minaa-webapp.git}, where users can access the source code and detailed documentation. An embedded tutorial video demonstrates how to navigate the interface, configure parameters, and interpret results, making the app accessible to users with varying technical expertise.

\section{Future work}
The \texttt{MiNAA-WebApp} will evolve to meet the needs of researchers in biological network analysis. Future updates will enhance scalability, introduce multi-network alignment, expand biological similarity metrics, improve visualization, and increase interoperability with bioinformatics tools.
We plan to optimize computational performance for handling larger networks, using memory-efficient algorithms and parallel computations. Multi-network alignment will enable simultaneous comparisons of multiple microbial communities. We will also broaden biological similarity metrics to include functional, phylogenetic, and metabolic correlations, and add topological features like centrality measures, clustering coefficients, and modular structures to refine network characterization and alignment accuracy.
These updates will maintain the \texttt{MiNAA-WebApp} as a versatile tool for a wide range of research applications.

\subsection*{Acknowledgements}
This work was supported by the National Science Foundation [DEB-2144367 to CSL].

\bibliographystyle{plain}
\bibliography{references}

\end{document}